# Counting Perfect Matchings in Dense Graphs Is Hard

Nicolas El Maalouly    Yanheng Wang

ETH Zürich, Switzerland

*October 26, 2022*

**Abstract**

We show that the problem of counting perfect matchings remains #P-complete even if we restrict the input to very dense graphs, proving the conjecture in [5]. Here "dense graphs" refer to bipartite graphs of bipartite independence number $\beta \leqslant 2$, or general graphs of independence number $\alpha \leqslant 2$. Our proof is by reduction from counting perfect matchings in bipartite graphs, via elementary linear algebra tricks and graph constructions.

## 1  Notations and Technical Lemmas

First let us fix some notations.

- Problem $\mathsf{Permanent}(\mathcal{G})$: How many perfect matchings are there for a given graph $G \in \mathcal{G}$?
- $\mathcal{B}$: all bipartite graphs.
- $\mathcal{B}'$: all complete bipartite graphs with potential parallel edges.
- $\mathcal{B}''$: all $(n-3)$-regular bipartite graphs.
- $\mathcal{C}$: all graphs with independence number at most 2.
- $f(n) := \begin{cases} (n-1)!! & n \text{ even} \\ 0 & n \text{ odd} \end{cases}$ counts the number of perfect matchings in $K_n$.

Towards our hardness results, we will study two special classes of square matrices. A main technical tool is the following theorem from linear algebra.

**Theorem 1.** *(Schur product theorem)* If matrices $M_1, M_2$ are positive definite, then their entry-wise product $M_1 \circ M_2$ is also positive definite.

**Lemma 2.** The matrix
$$A_n = \begin{pmatrix} 0! & 1! & \cdots & n! \\ 1! & 2! & \cdots & (n+1)! \\ \vdots & \vdots & \ddots & \vdots \\ n! & (n+1)! & \cdots & (2n)! \end{pmatrix}$$
is positive definite for all $n \in \mathbb{N}$.

*Proof.* Let us index the rows and columns from 0 to $n$. We pull out a factor of $i!$ from each row $i$ and a factor of $j!$ from each column $j$. This leaves us with a matrix $A'$ where $a'_{ij} = \frac{(i+j)!}{i! \cdot j!} = \binom{i+j}{j}$. It is a so-called *symmetric Pascal matrix* [4]. Observe that
$$\binom{i+j}{j} = \sum_{k=0}^{n} \binom{i}{k}\binom{j}{j-k} = \sum_{k=0}^{n} \binom{i}{k}\binom{j}{k}$$
where the first equality follows from a thought experiment. Suppose we are electing $j$ leaders out from $i + j$ candidates. To implement this, divide the candidates into two fixed groups of sizes $i$ and $j$, respectively. We elect $k$ leaders from the first group and the remaining $j - k$ from the second group, for a varying parameter $k$.

Given the identity, it follows that
$$A' = L\,L^{\mathrm{T}} \quad \text{where} \quad \ell_{ik} := \binom{i}{k}.$$

Clearly $L$ is a lower triangular matrix with an all-1 diagonal, which is invertible. So the matrix $A' = L L^{\mathrm{T}}$ (and hence $A_n$) is positive definite. □

**Lemma 3.** The matrices
$$B_n := \begin{pmatrix} f(0) & f(2) & \cdots & f(n) \\ f(2) & f(4) & \cdots & f(n+2) \\ \vdots & \vdots & \ddots & \vdots \\ f(n) & f(n+2) & \cdots & f(2n) \end{pmatrix} \text{ and } C_n := \begin{pmatrix} f(2) & f(4) & \cdots & f(n) \\ f(4) & f(6) & \cdots & f(n+2) \\ \vdots & \vdots & \ddots & \vdots \\ f(n) & f(n+2) & \cdots & f(2n-2) \end{pmatrix}$$
are positive definite for all $n \in 2\mathbb{N}$.

*Proof.* Note that $C_n$ is essentially $B_n$ without the first row and the last column. So it suffices to show $B_n$ is positive definite, and the property automatically transfers to $C_n$.

Let us index the rows and columns of $B_n$ from 0 to $n/2$, thus $b_{ij} = f(2(i+j))$. Recall that for any $t \in \mathbb{N}$, we have
$$f(2t) = (2t-1)!! = \frac{(2t)!}{(2t)!!} = \frac{(2t)!}{2^t t!} = \frac{t!}{2^t} \cdot \binom{2t}{t}.$$

This motivates us to split $B_n = U \circ V$, where
$$u_{ij} := \frac{(i+j)!}{2^{i+j}} \text{ and } v_{ij} := \binom{2(i+j)}{i+j}.$$

Observe that $U$ becomes the matrix $A_{n/2}$ in Lemma 2 if we multiply $2^i$ to each row $i$ and $2^j$ to each column $j$. Hence $U$ is positive definite.

Next we argue that $V$ is positive definite as well. First, we have identity
$$\binom{2(i+j)}{i+j} = \sum_{k=-n}^{n} \binom{2i}{i-k}\binom{2j}{j+k} = \sum_{k=-n}^{n} \binom{2i}{i-k}\binom{2j}{j-k}$$
due to the same thought experiment as before. Second, using the symmetry $\binom{2i}{i-k} = \binom{2i}{i+k}$ and $\binom{2j}{j-k} = \binom{2j}{j+k}$, we see that the terms for $k$ and $-k$ have the same value. So we may conclude
$$\binom{2(i+j)}{i+j} = \binom{2i}{i}\binom{2j}{j} + 2 \cdot \sum_{k=1}^{n} \binom{2i}{i-k}\binom{2j}{j-k},$$
and consequently
$$V = L \operatorname{diag}(1, 2, \ldots, 2) L^{\mathrm{T}} \quad \text{where} \quad \ell_{ik} := \binom{2i}{i-k}.$$

Because $L$ is a lower triangular matrix with an all-1 diagonal, $V$ must be positive definite. Finally, by Theorem 1 we see $B_n = U \circ V$ is positive definite. □

*Remark.* The matrix $V$ in the proof, among many other Hankel matrices starring binomial coefficients, are studied in combinatorics. See for example [1] and [2].

## 2 Hardness for bounded $\beta$

Now we are ready to prove our hardness results. The method is inspired by Okamoto, Uehara and Uno [6].

**Theorem 4.** Permanent($\mathcal{B}$) reduces to Permanent($\mathcal{B}'$). As a result, the latter is #P-complete.

*Proof.* Given a bipartite $G \in \mathcal{B}$ with $n$ vertices on each side, we construct a graph $G_i \in \mathcal{B}'$ for each $i = 0, \ldots, n$ as follows:
- add $i$ vertices to the left part and $i$ vertices to the right part;
- then add an edge for each left-right vertex pair, even if they were connected in $G$.

Let $p_i$ be the number of perfect matchings in $G_i$, which is assumed efficiently computable.

Denote by $m_j$ the number of matchings $M \subseteq E(G) : |M| = j$. Every such $M$ extends to exactly $(n+i-j)!$ perfect matchings $M' \subseteq E(G_i)$ with $M' \cap E(G) = M$. Clearly different $j, M$ contribute distinct $M'$, and they cover all possible perfect matchings. Hence $p_i = \sum_{j=0}^{n} (n+i-j)! \cdot m_j$. Writing in matrix form, we have $(p_0, \ldots, p_n) = (m_n, \ldots, m_0) A_n$, where $A_n$ is exactly the invertible matrix in Lemma 2. Hence we could recover $(m_n, \ldots, m_0)$, in particular $m_n$, from vector $(p_0, \ldots, p_n)$. □

*Remark.* Our $\mathcal{B}'$ allows parallel edges, which is somewhat undesirable. Better reduction exists in the literature. Dagum and Luby [3] gave a purely combinatorial reduction to Permanent($\mathcal{B}''$). Since any $G \in \mathcal{B}''$ has $\beta(G) \leqslant 3$, their result has similar philosophical implication.

## 3 Hardness for bounded $\alpha$

**Theorem 5.** Permanent($\mathcal{B}$) reduces to Permanent($\mathcal{C}$). As a result, the latter is #P-complete.

*Proof.* Given a bipartite $G \in \mathcal{B}$ with $n$ vertices on each side, we construct a graph $G_i \in \mathcal{C}$ for each $i = 0, \ldots, n$ as follows:
- add $i$ vertices to the left part and $i$ vertices to the right part;
- then connect every pair of vertices in the left (resp. right) part.

By a mirror argument to Theorem 4, we establish a linear equation $p_i = \sum_{j=0}^{n} f^2(n+i-j) \cdot m_j$. Writing in matrix form, we have $(p_0, \ldots, p_n) = (m_n, \ldots, m_0) Q$ where

$$Q := \begin{pmatrix} f^2(0) & f^2(1) & \cdots & f^2(n) \\ f^2(1) & f^2(2) & \cdots & f^2(n+1) \\ \vdots & \vdots & \ddots & \vdots \\ f^2(n) & f^2(n+1) & \cdots & f^2(2n) \end{pmatrix}.$$

It remains to prove that $Q$ is invertible, so that we could recover $(m_n, \ldots, m_0)$, in particular $m_n$, from vector $(p_0, \ldots, p_n)$.

As before, we index the rows and columns from 0 to $n$. Observe that $Q$ has a "checkerboard" pattern since $q_{ij} = 0$ iff $i + j$ is odd. To clean up the picture, we lift even rows to the top, and then push even columns to the left. When $n$ is even we derive

$$Q' = \begin{pmatrix} B_n \circ B_n & \mathbf{0} \\ \mathbf{0} & C_n \circ C_n \end{pmatrix},$$

and similarly, when $n$ is odd we derive

$$Q' = \begin{pmatrix} B_{n-1} \circ B_{n-1} & \mathbf{0} \\ \mathbf{0} & C_{n+1} \circ C_{n+1} \end{pmatrix}.$$

By Theorem 1, both the top-left and bottom-right blocks are positive definite, hence invertible. Therefore $\det(Q') \neq 0$, showing the invertibility of $Q'$ (and thus also $Q$). □